# Flame Synthesis of Graphene Films in Open Environments

*Nasir K. Memon[1], Stephen D. Tse[1*], Jafar F. Al-Sharab[2], Hisato Yamaguchi[2], Alem-Mar B. Goncalves[3], Bernard H. Kear[2], Yogesh Jaluria[1], Eva Y. Andrei[3], and Manish Chhowalla[2]*

[1]Department of Mechanical and Aerospace Engineering

[2]Department of Materials Science and Engineering

[3]Department of Physics and Astronomy

Rutgers University

Piscataway, NJ 08854

* Corresponding author.  sdytse@rci.rutgers.edu

ABSTRACT

Few-layer graphene (FLG) is grown on copper and nickel substrates at high rates using a novel flame synthesis method in open-atmosphere environments. Transmittance and resistance properties of the transferred films are similar to those grown by other methods, but the concentration of oxygen, as assessed by XPS, is actually less than that for CVD-grown graphene under near vacuum conditions. The method involves utilizing a multi-element inverse-diffusion-flame burner, where post-flame species and temperatures are radially-uniform upon deposition at a substrate. Advantages of the flame synthesis method are scalability for large-area surface coverage, increased growth rates, high purity and yield, continuous processing, and reduced costs due to efficient use of fuel as both heat source and reagent. Additionally, by adjusting local growth conditions, other carbon nanostructures (i.e. nanotubes) are readily synthesized.

## 1. Introduction

Graphene comprises a single layer of $sp^2$-bonded carbon atoms with remarkable physical, photonic, and electronic properties [1,2]. Both single-layer and few-layer graphene possess unique properties that afford a wide range of applications, including high frequency transistors [3] and transparent electrodes [4]. Ultimately, the future of graphene-based devices lies in developing production methods that are highly scalable, reliable, efficient, and economical. Mechanical exfoliation enabled the isolation of graphene and the discovery of its extraordinary electronic properties; however, this method is limited to producing graphene flakes due to its lack of scalability. Sublimation of Si from single-crystal silicon carbide (SiC) offers the advantage of direct synthesis of graphene on insulating surfaces [5,6]. Nevertheless, this method requires very-high temperatures, which has associated difficulties, and is presently constrained by high SiC wafer cost. Chemical vapor deposition (CVD) of graphene on transition metals such as nickel (Ni) [7,8] and copper (Cu) [9,10] shows the most potential for large-volume production of graphene. While still in its early stages, CVD-grown graphene has already demonstrated excellent device characteristics [11], including electron mobility of 7,350 $cm^2V^{-1}s^{-1}$ [12]. Nevertheless, growth of graphene over large areas remains challenging, due to the confinement necessary to operate at reduced pressures or suitable environments.

Flame synthesis has a demonstrated history of scalability and offers the potential for high-volume continuous production at reduced costs [13]. In utilizing globally-rich combustion, a portion of the hydrocarbon gas provides the requisite elevated temperatures, with the remaining fuel serving as the hydrocarbon reagent for carbon-based nanostructure growth, thereby constituting an efficient source of energy and hydrocarbon reactant. This aspect can be especially advantageous as the operating costs for producing advanced materials, particularly in the semiconductor industry, end up far exceeding the initial capital equipment costs.

Flame synthesis has been used successfully to grow various oxide nanostructures [14,15], single-wall [16] and multi-wall [17] carbon nanotubes (CNTs), sheet-like carbon particles [18], and amorphous carbon thin-films [19]. Recently, few-layer graphene has been synthesized with flames using alcohol as fuel on Ni substrates [20]. The process utilized two different burners, with the substrate situated within the interior region of the flame structure itself. Although the viability of flame synthesis to grow graphene was demonstrated, the process resulted in the formation of amorphous carbon impurities along with the graphene. Moreover, the configuration may not readily scalable for large-area graphene production. Flame synthesis of graphene on Cu has yet to be reported.

The unique synthesis configuration employed in this work is based on a multiple inverse-diffusion (non-premixed) flame burner, where the post-flame species are directed at a substrate to grow graphene; see Fig. 1. Each of the tiny diffusion flames is run in the inverse mode ("under-ventilated"), where for each flame, the oxidizer is in the center and fuel (e.g. methane) surrounds it. The net effect is that post-flame gases are largely comprised of pyrolysis species that have not passed through the oxidation zone. In fact, the reaction zone serves as a "getterer," such that the oxygen mole fraction can be reduced to $\sim 10^{-8}$ in the post-flame gases. Carbon formation processes are effectively separated from oxidation processes in inverse diffusion flames, which also tend to soot less than normal diffusion flames [21]. No soot is observed in our multiple-inverse diffusion flame setup, for the conditions examined. Moreover, the hydrocarbon species (rich in $C_n$ and CO), which serve as reagents for graphene growth, are generated in much greater quantities than that achievable in stable, self-sustained premixed flames. By using diffusion flames (burning stoichiometrically in the reaction zone), flame speed, flashback, and cellular instabilities related to premixed flames are avoided. Operation of a multi-element non-premixed flame burner has no scaling problems by allowing for stability at all burner diameters, where the issuing flow velocity can be independent of the burner diameter. Moreover, since many small diffusion flames are utilized, overall radially-flat profiles of temperature and chemical species are established downstream of the burner, ensuring uniform growth. Confinement in an inert environment or shielding with an inert co-flow or tube prevents an encompassing diffusion flame to develop. Finally, this flame synthesis configuration is well suited for carbon-based nanomaterial synthesis in open-atmosphere environments, affording large-area growth (e.g. by translating the burner and rasterizing) at high rates.

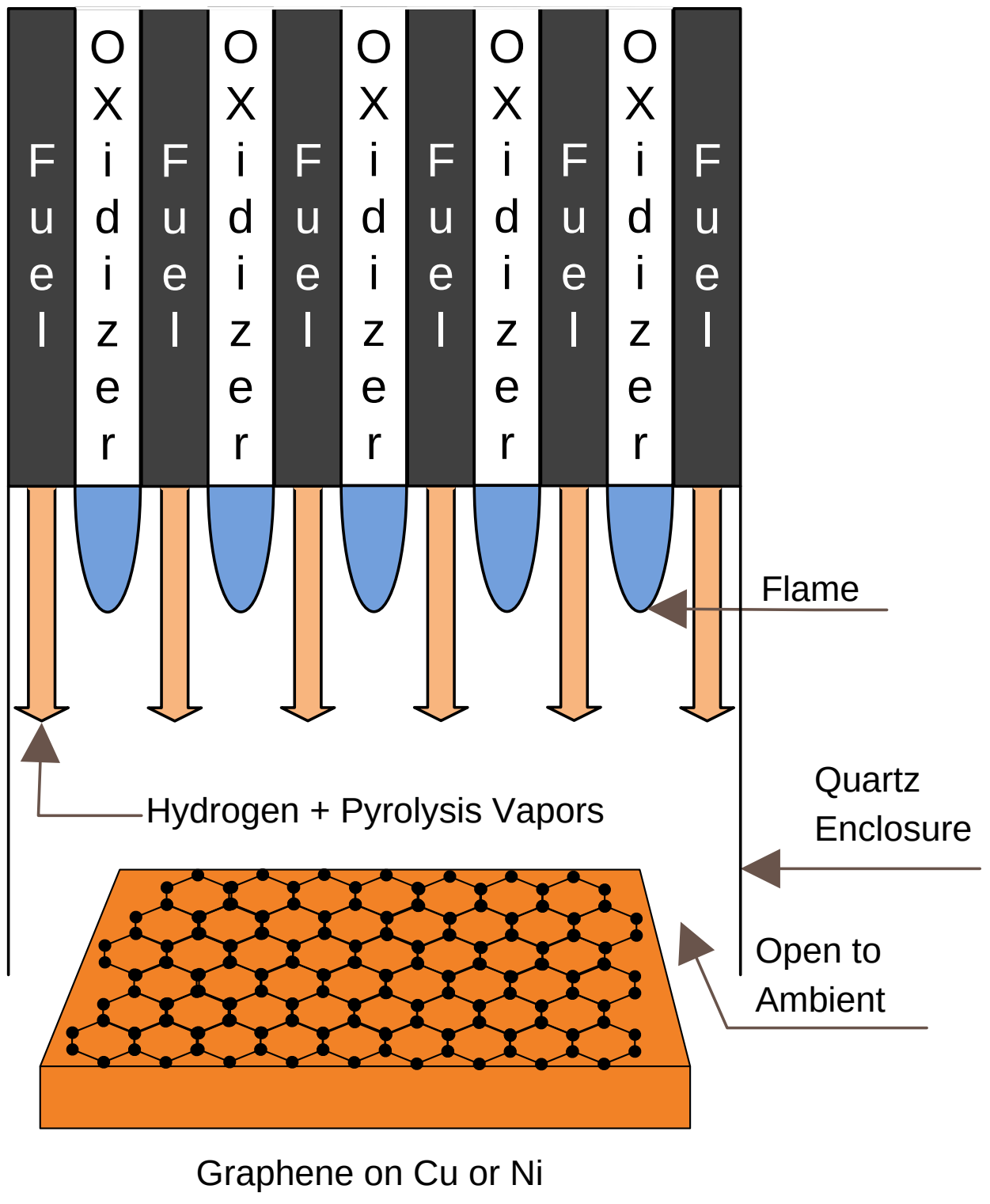

**Figure 1**. Multiple inverse-diffusion flames provide hydrogen and carbon-rich species suitable for growth of graphene and other carbon nanomaterials.

## 2. Experiment

Few-layer graphene (FLG) films are grown on 25 μm thick Cu and Ni foils (Alfa Aesar), placed downstream of our novel burner. A quartz cylinder encompasses the region of the multiple flames and substrate, preventing oxidizer permeation from the ambient and directing optimal gas-phase conditions (i.e. species and temperature) to the substrate. Note that the setup is open to atmospheric conditions. Prior to FLG film synthesis, the metal substrates are reduced in a hydrogen environment to remove any oxide layers. This treatment is accomplished using the same multiple inverse-diffusion flame burner running only hydrogen as fuel at a globally-rich equivalence ratio for 10 minutes. For FLG synthesis, $CH_4$ is introduced into the fuel (with a global equivalence ratio of ~3) for 5 and 10 min, for Ni and Cu substrates, respectively. A silica-coated 125 μm Pt/Pt-10%Rh thermocouple (S-type) measures the substrate temperature to be ~950$^{\circ}$C. The experiment is finalized by turning off the oxygen, which extinguishes the flame, while fuel and inert gases continue to flow, cooling the substrate to room temperature.

The films grown on Cu are transferred onto $SiO_2$/Si substrates for electrical and Raman analysis. The transfer is done by first spin-coating poly-methyl methacrylate (PMMA) on the graphene covering the Cu substrate. Since the thermofluid mechanics give rise to FLG being grown on both sides of the substrate, oxygen plasma is used to remove the graphene film from one side. The PMMA-coated graphene on Cu substrate is then immersed in a ferric chloride ($FeCl_3$) solution (23%wt) to etch away the copper. The free-floating PMMA coated graphene is then carefully placed on the $SiO_2$/Si substrate, and the PMMA is removed in hot acetone. The final sample is rinsed with isopropanol, and dried with $N_2$.

The FLG is characterized using Raman spectroscopy (Renishaw 1000, laser excitation 514.5 nm), atomic force microscopy (AFM, Digital Instruments Nanoscope II), X-ray photoelectron spectroscopy (XPS, Thermo Scientific K-Alpha), and transmission electron microscopy (TEM, JEOL 2010F). The CNTs are examined using scanning electron microscopy (SEM, Zeiss Sigma 8100).

## 3. Results and Discussion

Synthesis of FLG has been demonstrated on a number of transition metals. Due to their low cost and acceptability in the semiconductor industry, copper and nickel are promising substrates for the growth of graphene. While a number of parameters such as pressure, temperature, and crystal structure influence the growth of graphene, the difference in carbon solubility of metals such Cu and Ni results in distinctive growth mechanisms [22]. From the binary phase diagram of Ni and C [23], at temperatures above 800°C, Ni and C form a metastable solid phase; upon cooling, the carbon diffuses out of the Ni to form graphene/graphite. Due to this growth mechanism on Ni, the number of graphene layers across the substrate remains difficult to control. In contrast, graphene formation on Cu occurs only on the surface due to the extremely-low solubility of carbon in Cu. Consequently, once the substrate is covered by graphene, the Cu surface is no longer accessible; and deposition of additional layers does not occur [9, 22]. Hence, Cu has proven to be an excellent substrate for the growth of monolayered graphene; however, growing multiple layers has been found to be challenging.

### 3. 1 Flame Synthesis of FLG on Cu

A photograph of a flame-synthesized FLG film that has been subsequently transferred onto a 1cm × 1cm quartz substrate is shown in Fig. 2a. In Fig. 2b, an optical image shows a graphene flake along with the corresponding atomic force microscopy (AFM) image. The thickness of the graphene films on Cu is found typically to be on the order of 4nm from AFM height profiles, suggesting that the film consists of 8 to 10 monolayers of graphene.

Raman spectroscopy enables the identification of single to few-layer graphene [24], along with its quality. Typical Raman spectrum of FLG after transfer onto $SiO_2$/Si is shown in Fig. 2(c). Three peaks are noticeably present in the spectrum: (*i*) the D peak at 1351 $cm^{-1}$, which is due to the first-order zone boundary phonons and is used to determine the disorder present in the graphene; (*ii*) the G peak at 1580 $cm^{-1}$, which is related to the bond stretching of $sp^2$ bonded carbon atoms; and (*iii*) the 2D peak at ~2700 $cm^{-1}$, which is caused by the second-order zone boundary phonons. The ratio between the intensities of the G peak ($I_G$) and the 2D peak ($I_{2D}$) provides an estimate of the number of layers [8, 25], where, from Fig. 2(d), the values are found to range from 1.3 to 1.7. For mono and bi-layer graphene, this ratio is less than 1. If more than 2 layers are present, ratios ranging from 1.3 to 2.4 have been reported for FLG. Reina et al. [25] reported $I_G/I_{2d}$ ratio of 1.3 for 3 layers of graphene on Ni; and Robertson et al. [26] reported values of 1.8 to 2.4 for 5 to 10 layers of graphene on Cu. The full-width and half-maximum (FWHM) of our 2D peak is ~75 $cm^{-1}$, which is consistent with FLG grown at atmospheric pressure [10]. The Raman data should be used in conjunction with other characterization and verification techniques to corroborate the properties of FLG. Transmittance can be used to assess the number of graphene layers, where the opacity of monolayer graphene is estimated to be 2.3% [27]. From Fig. 3a, the transmittance of our FLG films at 550nm is 86%, which correlates to ~6 layers. Combining our results from Raman, AFM, and transmittance, we estimate that 5 to 8 layers of graphene are grown uniformly across the Cu substrate using our flame-synthesis technique.

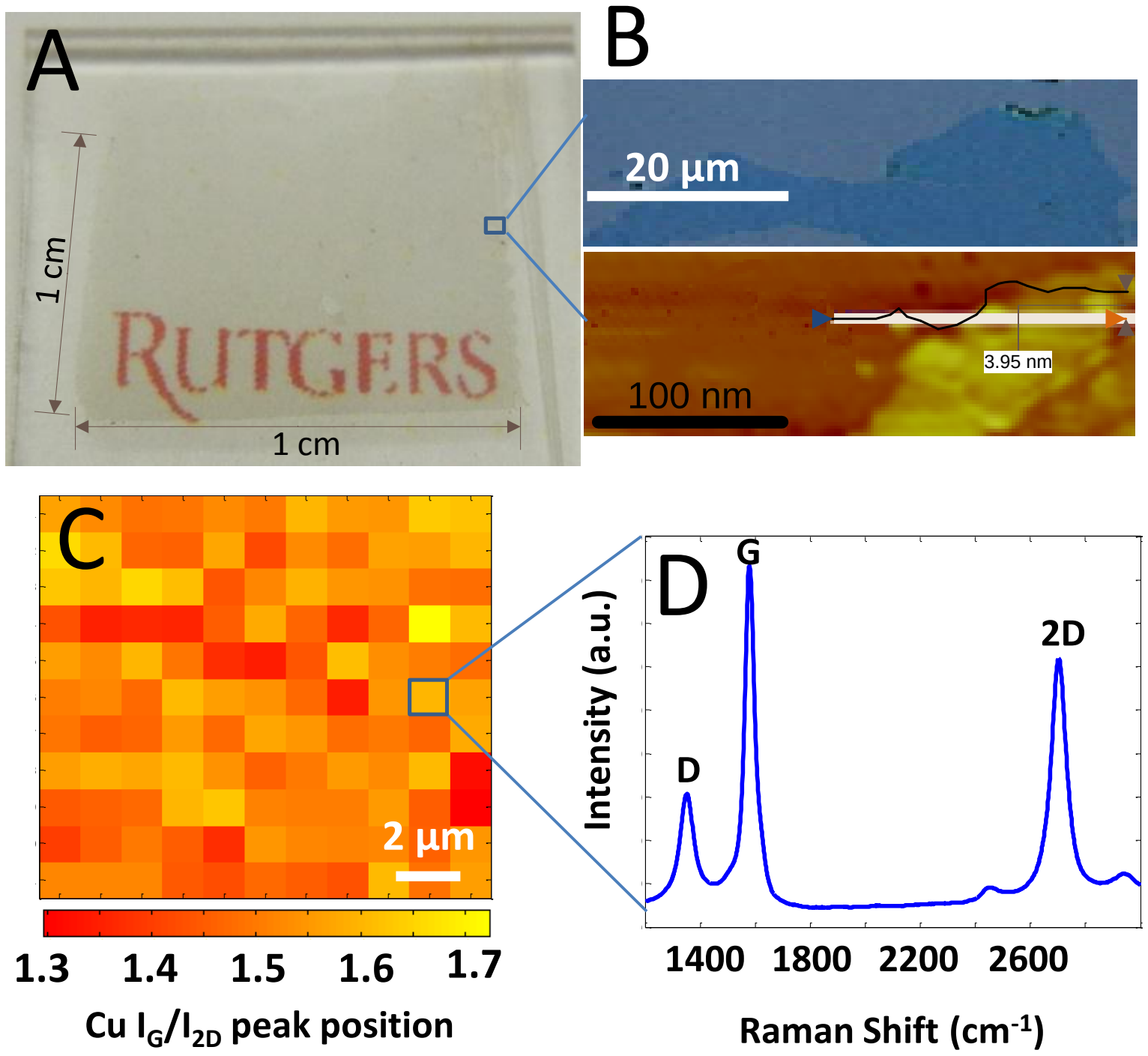

**Figure 2.** Few-layer graphene (FLG) film grown by flame synthesis on Cu. (a) Photograph of a 1cm x 1cm film transferred onto quartz. (b) Optical microscope image of the FLG transferred onto a Si substrate with a 300nm oxide layer and the corresponding AFM image and height profile of the FLG transferred on $SiO_2$/Si substrate. (c) Two-dimensional mapping of the Raman $I_G/I_{2D}$ over a 12μm x 12 μm area. (d) Raman spectrum of the FLG on $SiO_2$/Si.

The $I_D/I_G$ ratio observed in our sample is around ~0.35, which is comparable to measurements of FLG grown using other methods [10]. The measured disorder in our FLG likely arises from the sheets being composed of sub-micron domains. Using the four probe method, the sheet resistance of the FLG is calculated to be 40kΩ/sq at 86% transmittance value, which is considerably higher than CVD-grown graphene. Similarly, the high sheet resistance may be attributed to the small domain size of the graphene. In CVD growth, the characteristic domain size of graphene has been increased by lowering the global flux of methane [28].

However, the fundamental mechanism for this trend is not clear, as there are many effects intertwined; and additional parameter dependencies need to be explored to isolate the controlling mechanism dictating domain size. We are currently investigating the effect of methane flux, as well as other parameters, on enlarging domain size for flame-synthesized graphene on Cu.

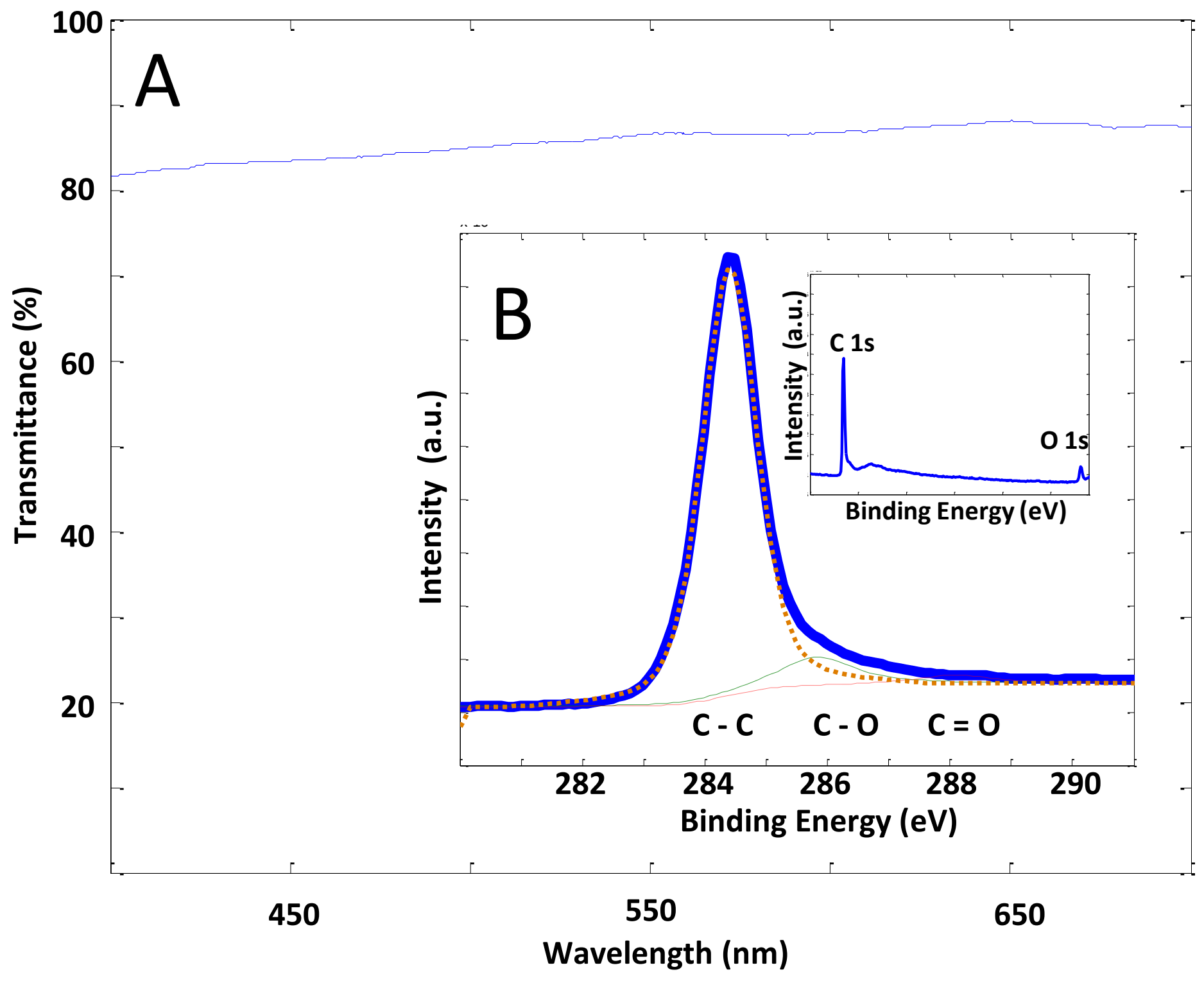

**Figure 3**. (a) UV-vis spectrum of the FLG transferred onto a quartz substrate. (b) XPS C 1s spectrum of the FLG shows that the oxygen contamination is minimal and is comparable to CVD grown graphene. The inset shows the XPS spectra of the film.

A product of hydrogen and hydrocarbon combustion with oxygen is $H_2O$, which at high temperatures can result in oxygen doping of graphene. However, with abundant $H_2$ present in the post-flame species, and at relatively "low" growth temperatures (~950ºC), such oxidation reactions are minimized. The X-ray photoelectron spectroscopy (XPS) spectrum of the C 1s peak, where the main peak at 284.4eV indicates that most of the atoms are in the $sp^2$ C state, is shown in Fig. 3b. Less than 10% of oxygen incorporation (e.g. CO) is visible from the XPS figure. Surprisingly, the amount of oxygen-bonded species in our open-ambient flame synthesis process is actually lower than that for CVD-grown graphene under near vacuum conditions [11]. Consequently, the conditions for our flame synthesis, where $H_2O$ oxidation is minimized, $O_2$ is "gettered" in the reaction zone, and open-atmosphere processing is afforded, are advantageous for scaled growth of graphene over large areas (e.g. over existing structures). Note that the O 1s peak, as seen in the inset of Fig. 3b, is due to oxygen or water absorbed on the surface and is even present in pristine graphene [29].

The effects of $CH_4$:$H_2$ ratio and temperature are examined in the growth of FLG on Cu. For the standard case, the $CH_4$:$H_2$ ratio is kept at 1:10, and similar results are observed when this ratio is varied from 1:5 to 1:20. However, when the ratio is below 1:40, no growth of FLG is observed on the substrate. This result is contrary to that reported using atmospheric-pressure CVD [10], where at lower $CH_4$:$H_2$ ratios, monolayer graphene is synthesized. In flame synthesis, temperature is a critical factor in the growth of uniform FLG. At lower gas-phase temperatures, the typical Raman spectrum features resemble those of nanocrystalline graphite [30], where a much higher D-peak exists and the intensity ratio between the G peak and 2D peak increases, as shown in Fig. 4. Upon further reducing the gas-phase temperature, the 2D peak disappears. However, a G peak is still observed, indicating the presence of activated carbon-based materials on the copper [31]. The reason for different carbon-based growth on Cu is perhaps due to the presence of other gaseous carbonaceous species, such as CO and $C_n$, in the post-flame environment. These species can readily decompose at lower temperatures to form carbon materials that are stable at lower temperatures. In another work [19] that attempts to grow graphene on copper using flames, a thin carbon film is synthesized with large amounts of

sp$^3$ bonding. This characteristic of the thin film was attributed to the low deposition temperatures of 550°C to 700°C.

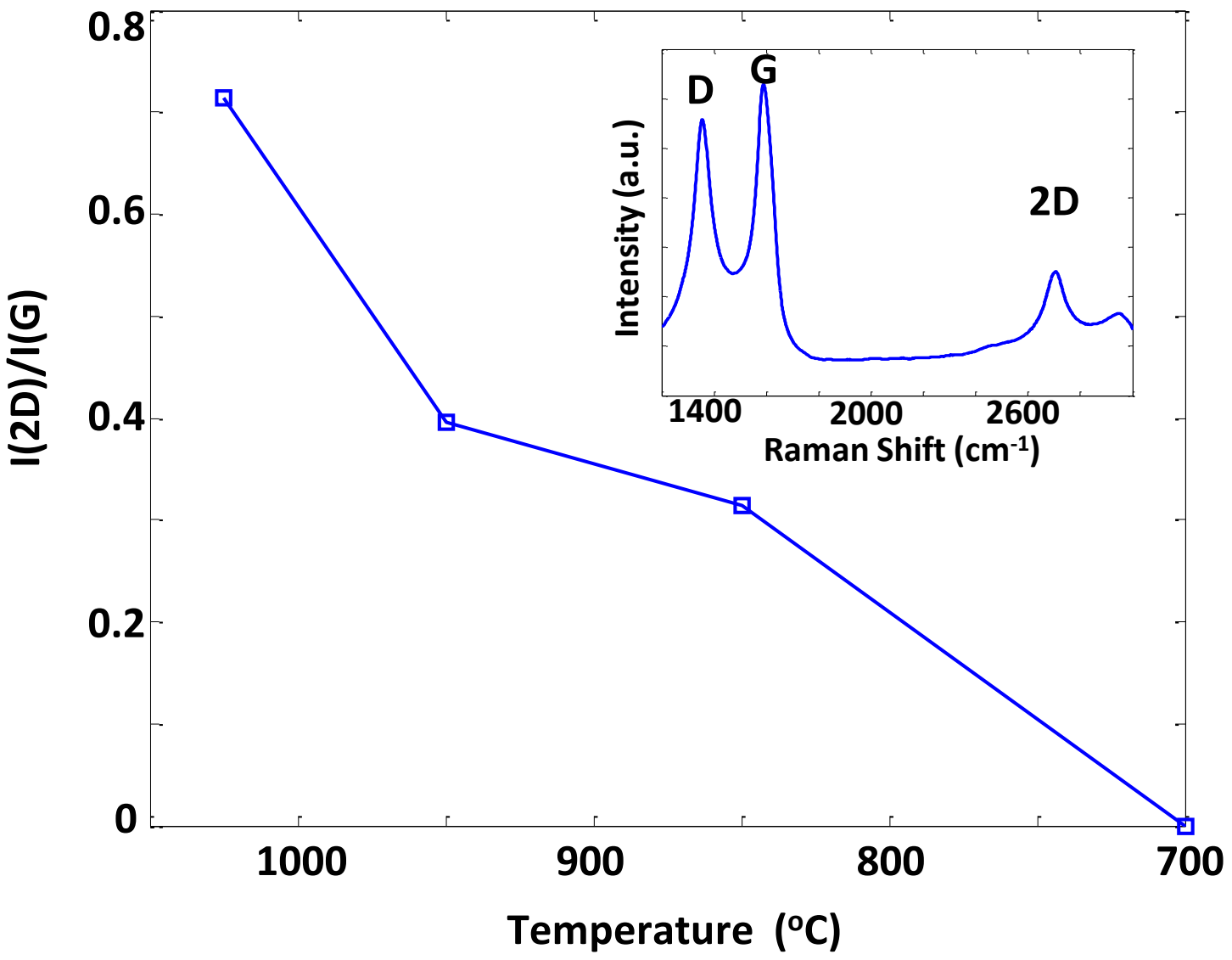

**Figure 4**. Analysis of the influence of temperature on the growth of FLG on Cu, showing variation in Raman $I_{2D}/I_G$ as a function of gas phase temperature. The inset shows atypical Raman spectrum observed at lower temperatures.

## 3.2 Flame Synthesis of FLG on Ni

A typical Raman spectrum of FLG grown on 25μm thick Ni foil is shown in Fig. 5a. The number of layers of graphene on nickel is estimated based on the location of the 2D peak [8]. With the 2D peak at 2720 cm$^{-1}$, this shift corresponds with 5 to 10 layers of graphene. The G peak is at 1583 cm$^{-1}$, and a typical $I_D/I_G$ ratio is 0.1, which is lower than that for the FLG grown on Cu. This result can be attributed to the different growth mechanism of graphene on Ni compared with that on Cu. Ni has higher carbon solubility, so the growth of graphene occurs due to carbon segregation or precipitation. This growth mechanism on Ni should be unaffected by the high carbon flux encountered in the flame. However, in the case of Cu, high carbon flux may lead to smaller graphene domain size, and hence more measured disorder. A HRTEM image and the corresponding diffraction pattern are shown in Fig. 5b. The hexagonal symmetry of multiple graphene layers can be inferred from the diffraction pattern, although specific stacking order of the layers requires additional analysis. A magnified image of the well-ordered graphitic lattice is shown in the inset of Fig. 5b.

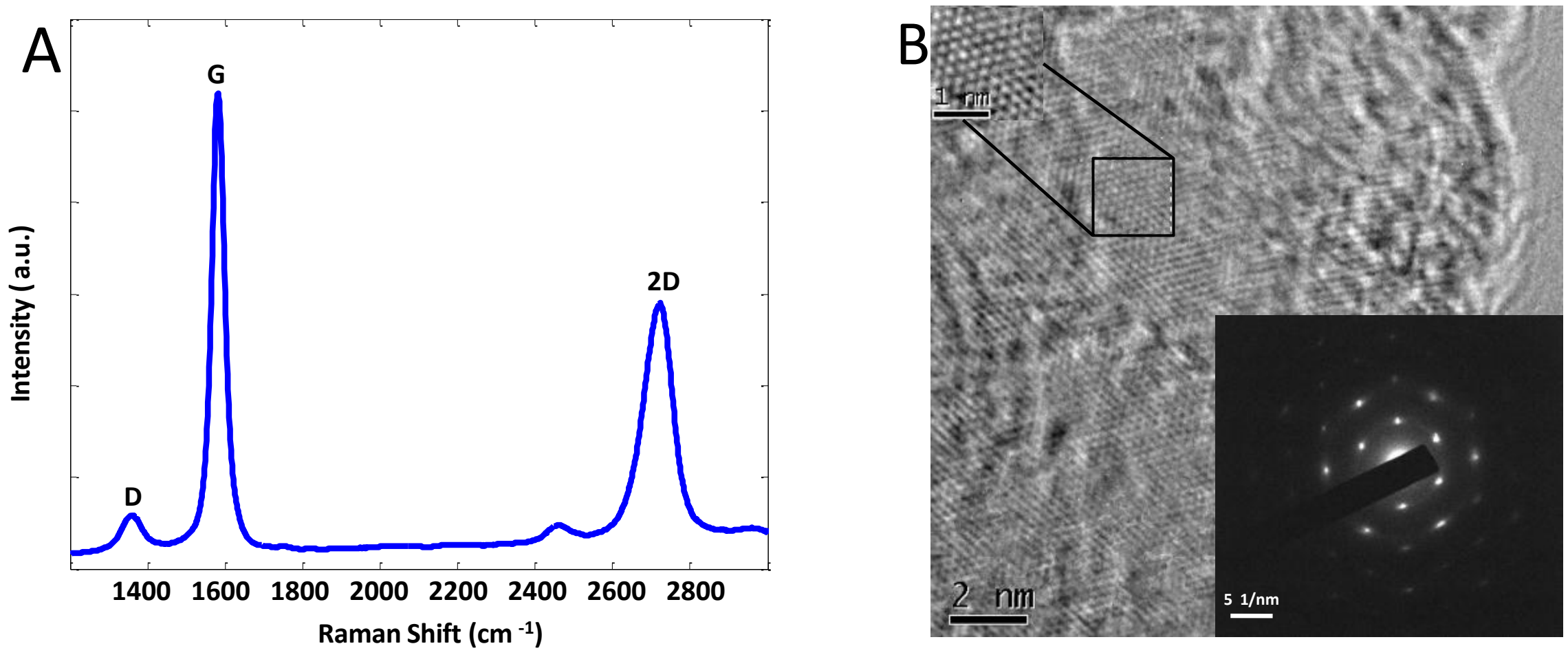

**Figure 5**. Few-layer graphene film grown by flame synthesis on Ni. a) Raman spectrum of the FLG on Ni. b) HRTEM image of the FLG. The bottom right inset shows the electron diffraction pattern of the graphene sheet, illustrating the well-defined crystalline structure. The top left inset shows resolution magnified image of the graphitic lattice.

Graphene growth on nickel depends on a number of parameters, such as metal substrate thickness, hydrocarbon to hydrogen ratio, growth time, and temperature. The dependence of graphene growth on gas-phase temperature and methane concentration is illustrated in Fig. 6. The gas-phase temperature directly affects the gas-phase chemistry as well as the rate of hydrocarbon decomposition on the Ni surface, which further affects the diffusion rate of carbon atoms into Ni [32]. Hence, at lower gas-phase temperatures, fewer carbon atoms diffuse into Ni, leading to the growth of fewer layers of graphene upon cooling. This effect is evident from optical microscopy of the as-synthesized graphene on Ni. At lower temperatures (i.e. 850ºC),

Fig. 6a shows lighter regions corresponding to FLG; and the color contrast demonstrates that the growth of FLG on Ni is not uniform. With increased temperature (i.e. 950ºC), Fig. 6b shows that the Ni foil becomes uniformly dark, indicating the presence of more than 10 layers of graphene. The dependence of graphene growth on the ratio of methane to hydrogen is shown in Fig. 6c. When this ratio is lowered to 1:20, the Raman position of the 2D peak (~2700 $cm^{-1}$) indicates that fewer than 5 layers of graphene are grown. Interestingly, such growth is similar to the 850ºC temperature growth illustrated in Fig. 6a, where the growth is non-uniform across the substrate.

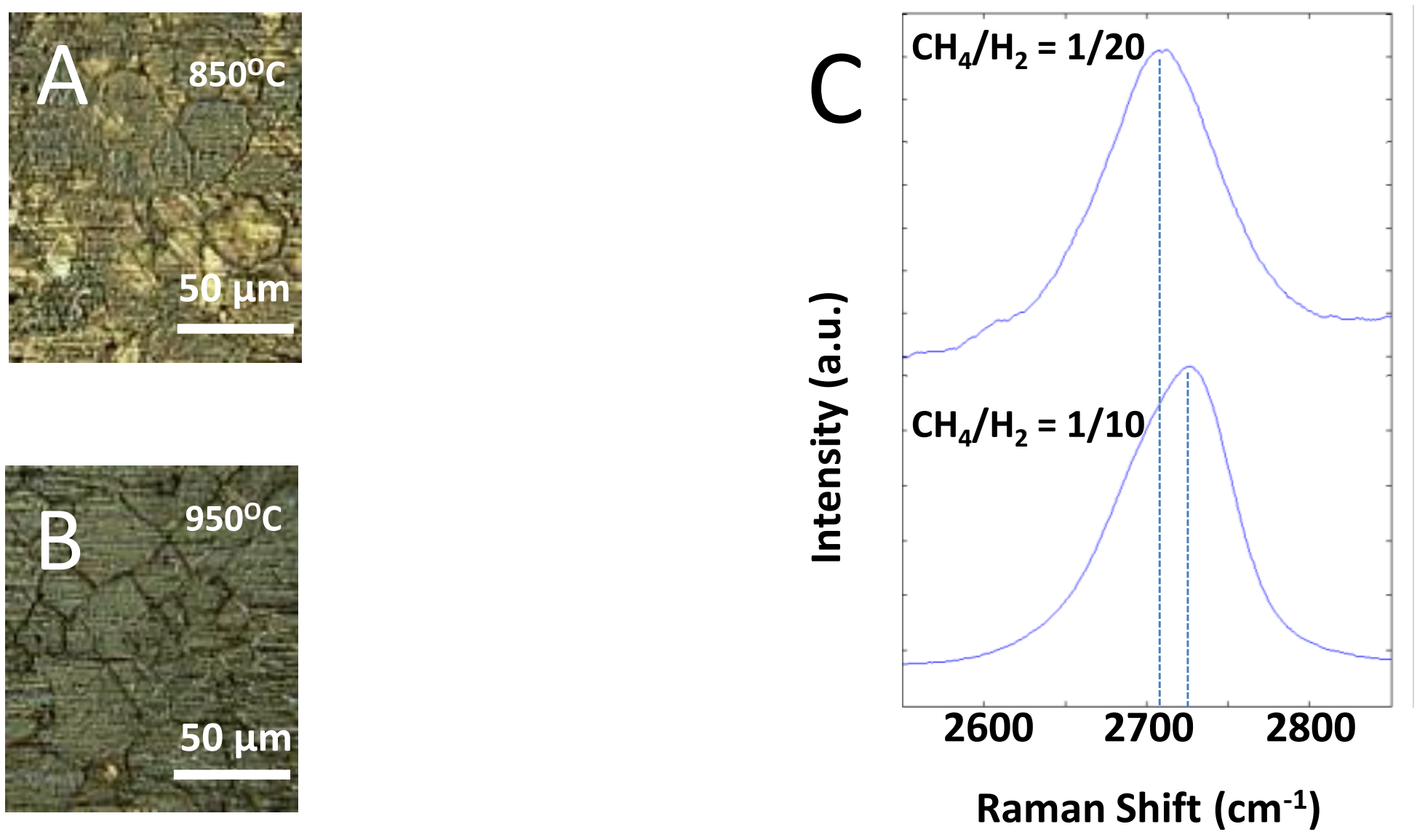

**Figure 6**. Investigation of the temperature and methane concentration on the growth of FLG on Ni. a) Temperature = 850°C, $CH_4$:$H_2$ = 1:10. b) Temperature = 950°C, $CH_4$:$H_2$ = 1:10. c) Raman spectra showing the 2D peak at a constant temperature of 950°C.

For the gas-phase synthesis conditions examined, the growth of FLG on Ni results in lower disorder, as assessed by Raman, when compared to the growth on Cu. However, the growth is less uniform and comprises more layers (>10), due to the different growth mechanism of graphene on Ni, compared to growth on Cu. The Raman mappings of FLG grown on Cu and Ni, respectively, at a $CH_4$:$H_2$ ratio of 1:10, are compared in Fig. 7. For Cu, methane is introduced for 10 minutes at a temperature of 950°C; while for Ni, methane is introduced for 5 minutes at a temperature of 850°C. In Fig. 7a, for Cu, the 2D peak is always at or below 2700 $cm^{-1}$, which is consistent with FLG. On the other hand, in Fig. 7b, for Ni, the 2D peak reaches a value of 2727 $cm^{-1}$, indicating the presence of more than 10 layers. As such, the growth of graphene on Cu is self-limiting to few-layers for our flame synthesis system at atmospheric conditions.

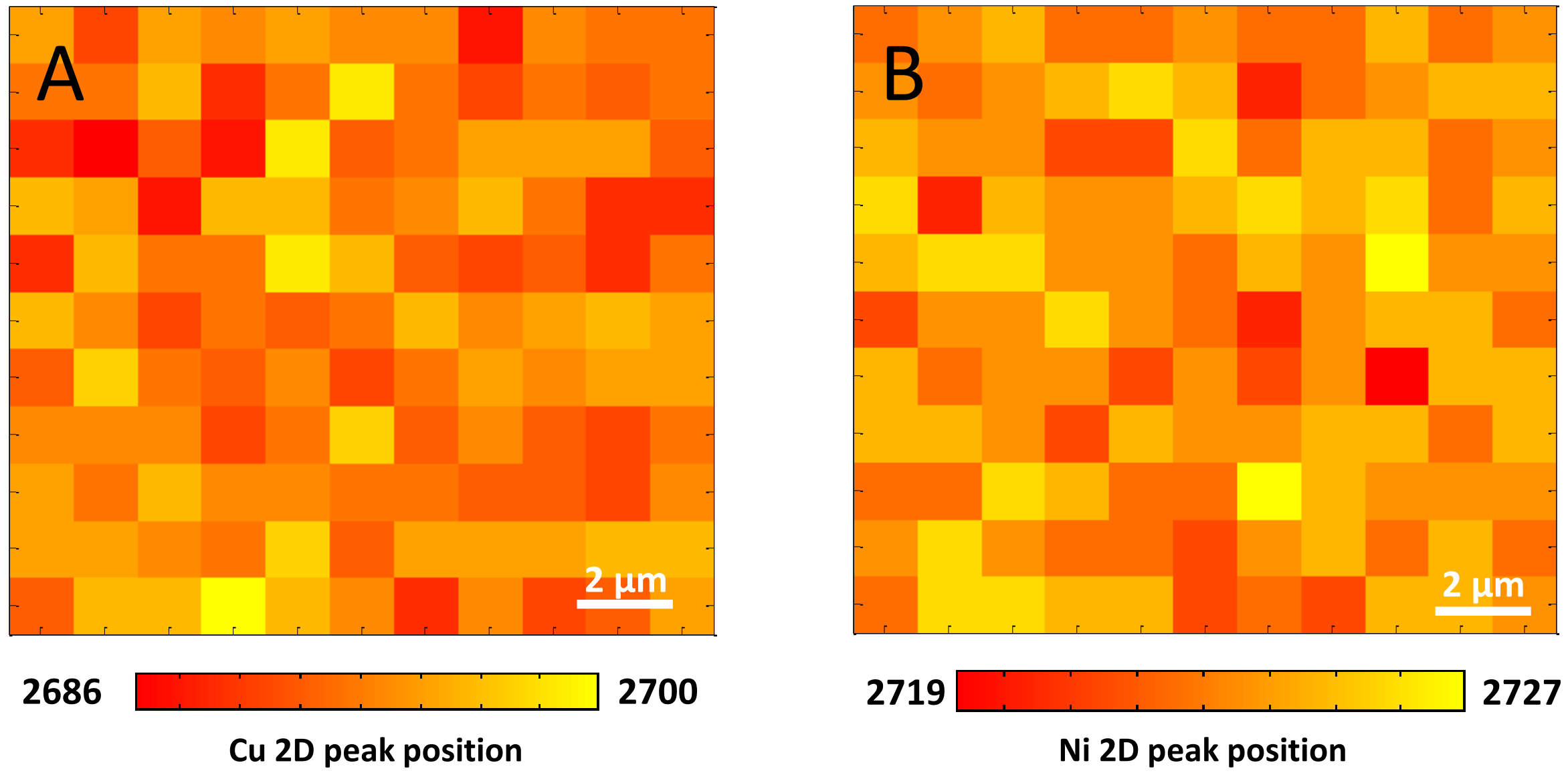

**Figure 7**. Raman mappings of the 2D peak over a 12 µm x 12 µm region at a constant $CH_4$:$H_2$ ratio of 1:10. a) Raman mapping for Cu, illustrating that the growth of graphene is self-limiting to a few layers. b) Raman mapping for Ni, showing regions that correlate to more than 10 layers.

## 3.3 Flame Synthesis of CNTs

By adjusting the conditions (e.g. temperature), our multiple inverse diffusion flame burner can readily synthesize other carbon-based nanomaterials such as CNTs and fullerenes. For example, for CNT growth, ethylene is used as the fuel source, with a Ni/Ti alloy substrate placed in the post-flame region. Transition metals (e.g. Ni, Co, and Fe) and their alloys are well known to serve as catalysts for CNT growth. Under the right conditions, catalyst nanoparticles are formed, and carbon-based precursor species readily undergo dissociative adsorption and diffuse through the catalyst nanoparticles and grow into CNTs. Using our flame setup, no pretreatment of the substrate is needed; our single-step method induces catalyst nanoparticle formation [17, 33] along with subsequent CNT growth. An SEM image of the as-grown CNTs is shown in Fig. 8. With the temperature and chemical species concentrations in the post-flame gases radially flat, uniform synthesis of CNTs is possible. Additionally, for fundamental study, the axial gradients are moderate, so that flame conditions can be parametrically examined to establish, with precision control, local "universal" conditions (e.g. gas-phase temperature, substrate temperature, relevant species) that correlate with resultant CNT morphologies and growth rates. The transition from graphene to CNT growth, with respect to local conditions as well as spatial interfaces, is currently being investigated.

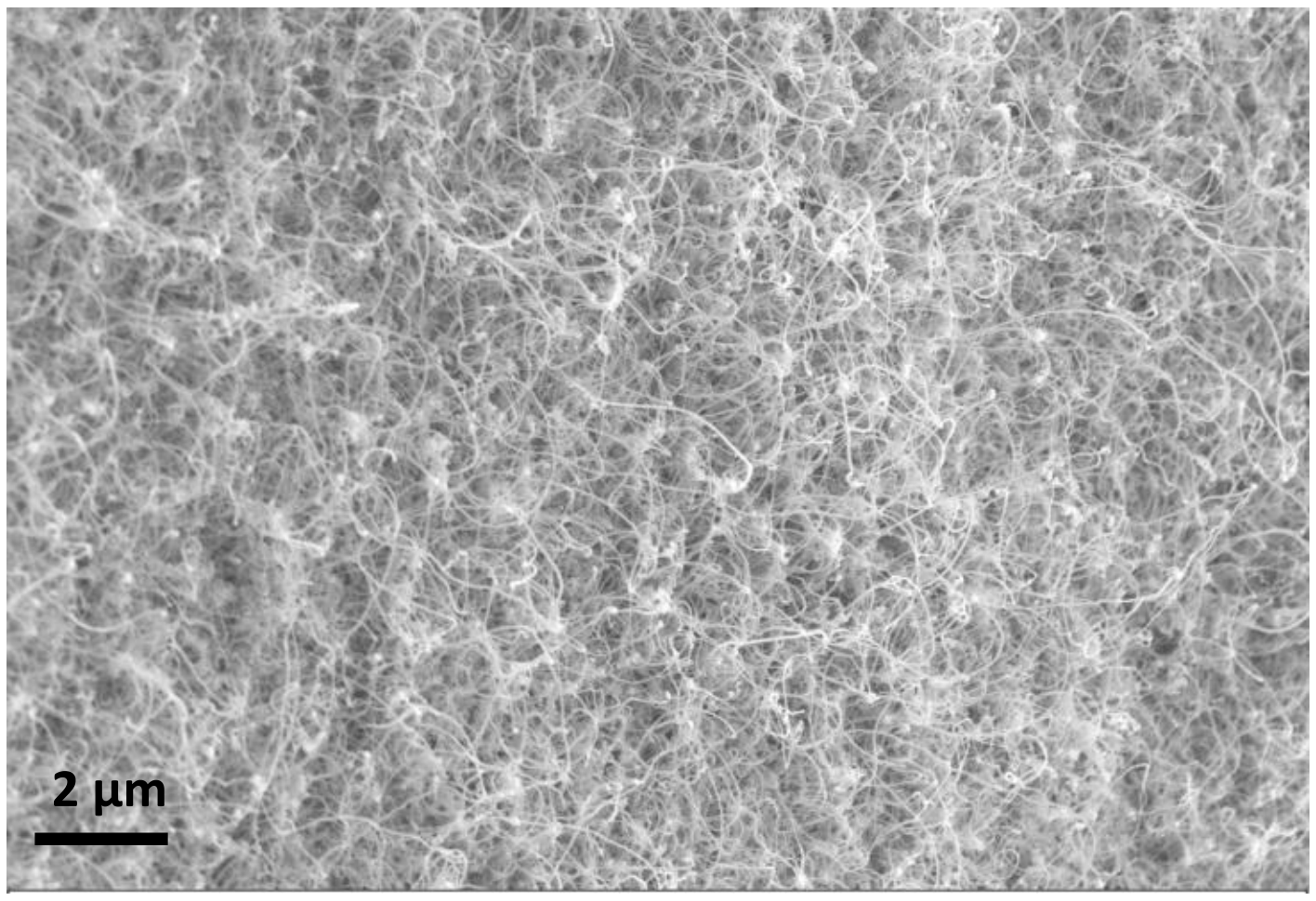

**Figure 8**. SEM image of CNTs grown on a Ni/Ti substrate.

## 4 Concluding Remarks

Flame synthesis utilizing a multiple-inverse diffusion flame burner is demonstrated in this work to be well-suited for processing carbon-based nanostructures. Under very rich fuel conditions, the configuration generates specific hydrocarbon species that can form graphene on a heated metal substrate. On Cu, 5 to 8 layers of graphene are grown uniformly across the substrate. Due to a different growth mechanism, Ni offers lower graphene disorder, but at a cost of more layers created. Nonetheless, the growth conditions have not been optimized in this study, and on-going parametric refinement should result in higher quality and fewer layers of graphene produced. The configuration allows for detailed probing of the local gas-phase temperature and relevant chemical species such that the fundamental growth mechanisms of graphene on various substrates can be identified.

The novel non-premixed flame synthesis process is expected to complement CVD-type processes in the growth of graphene and CNTs. Elevated gas-phase temperatures and flame chemistry provide the precursors for growth, making hydrocarbon (as well as doping precursor) decomposition more independent of substrate temperature, offering an additional degree of freedom in tailoring film characteristics. The encompassing quartz cylinder, which prevents oxidizer transport from the ambient, can also serve as a "reactor wall," whose cooling/heating rate can be tuned to optimize gas-phase chemistry and temperature reaching the substrate for ideal carbon-based growth. The present setup affords fast growth rates due to innately high flow rates of precursor species; control of temperature and reagent species profiles due to precise heating at the flame-front, along with self-gettering of oxygen; and reduced costs due to efficient use of fuel as both heat source and reagent. Growth is uniform because the configuration produces post-flame gases downstream that are quasi one-dimensional, i.e. radially-uniform in temperature and chemical species concentrations. Finally, the method is scalable and capable of continuous operation in an open-ambient environment, presenting the possibility of large-area processing.

## 5 Acknowledgements

This work was supported by the Army Research Office (Grant W911NF-08-1-0417), the Office of Naval Research (Grant N00014-08-1-1029), and the National Science Foundation (Grant 0903661, Nanotechnology for Clean Energy IGERT). Special thanks are due to Sylvie Rangan for her assistance with the XPS.